\documentstyle[aps,preprint]{revtex}

\begin{document}

\title{The Construction of Double-Ended Classical Trajectories}

\author{A. E. Cho\footnote{Present address:  Department of Supercomputer
Applications, Samsung Advanced Institute of Technology, P. O. Box 111,
Suwon, Korea 440-600 (a0escho@super.sait.samsung.co.kr)} and J. D. Doll}
\address{Department of Chemistry,
Brown University,
Providence, RI 02912}

\author{D. L. Freeman}
\address{Department of Chemistry,
University of Rhode Island,
Kingston, RI 02881}

\maketitle
\begin{abstract}
In the present paper we describe relaxation methods for constructing
double-ended classical trajectories.  We illustrate our approach with an
application to a model anharmonic system, the Henon-Heiles problem.
Trajectories for this model exhibit a number of interesting energy-time
relationships that appear to be of general use in characterizing the
dynamics.
\end{abstract}
\pacs{PACS numbers: }

\section {Introduction}
        As commonly formulated, problems in classical dynamics are
typically
``initial value'' in character.  That is, they involve the determination
of
the evolution of a set of coordinates and momenta given their values at a
specified starting time.  Assuming the relevant equations of motion are
known, this problem reduces to the numerical task of integrating these
equations forward for the required time period.  Although not without
technical challenge, such initial value problems are well posed and
amenable to study with established methods \cite {one}.

        In certain physical applications, however, ``double-ended''
formulations of
the dynamics prove more natural.  These include, for example, the
establishment of chemical transition states \cite{two}, the search for
mechanisms of complex reactions \cite {three}, and various semiclassical and
quantum formulations of dynamics \cite {four,five,six}.
It is more appropriate in
these applications to specify the initial and final positions for the
trajectory rather than initial positions and momenta.  Since double-ended
boundary conditions do not, in general, determine a unique path, there may
be zero, one, or many such trajectories for a specified transit time.

        Less developed than their initial value counterparts, the most common
technique for generating double-ended paths basically amounts to trajectory
``shooting.''  In this approach the required paths are established by varying
initial trajectory momenta in order to find all values that lead to the
desired final positions. Viable for relatively simple low-dimensional
problems, this procedure is cumbersome for general applications where
``aiming'' problems become severe.

     Relaxation methods for the construction of double-ended trajectories have
been discussed by Doll, Beck and Freeman \cite {seven}.  Related developments
have been presented more recently by Gillilan and Wilson \cite {eight}.  We
review and extend our original discussion in Section II below and describe
a brief application to model, anharmonic problem, the Henon-Heiles system,
in Section III.
\section{Formal Developments}

We adopt a one-dimensional notation in what follows as a matter of
convenience noting that the generalization to many-dimensional applications
is straightforward.

        We seek the classical path(s) for a one-dimensional conservative
system
with a known potential energy, $V(x)$, that begin at the point $x$ at time
zero
and end at the point $x'$ a time $t$ later.  To proceed it is convenient to
write the paths in terms of a dimensionless time, $u$, that ranges from zero
to unity.  In terms of this dimensionless time, an arbitrary path of the
type we seek can be written as  \cite {seven}
\begin{equation}
 x(u) = x + (x' - x) u + \sum^{\infty}_{k=1} a_k \sin (k\pi u).
\label{eq:1}
\end{equation}
The straight line portion of Eq. (1) builds the required boundary
conditions into the description while the remaining Fourier sum describes
all possible fluctuations about this reference path.  Formally an infinite
sum, the number of Fourier terms required in Eq. (1) to obtain convergence
in practice varies with application and generally increases with the
duration of the trajectory.

The desired classical path(s) are obtained by substituting the path ansatz
into the functional,
\begin{equation}
S([x(u)]) = t \int_o^1 du\left\{ \frac{m}{2t^2} \left( \frac{dx(u)}
{du}\right)^2 - V(x(u))\right\} ;
\label{eq:2}
\end{equation}
and requiring that the resulting action be stationary with respect to the
expansion coefficients,
\begin{equation}
\frac{\partial S}{\partial a_k} = 0.
\label{eq:3}
\end{equation}
This leads to the non-linear equations for the Fourier coefficients,
\begin{equation}
a_k = \frac{-t^2}{m\pi^2k^2}f_k (a).
\label{eq:4}
\end{equation}
Here $m$ is the particle's mass and $f_k$ is the $k$th Fourier sine
component of
the force along the path specified by the coefficients $\{a_k\}$,
\begin{equation}
f_k(a) = 2\int_0^1 du \sin (k\pi u)
\left(-\frac{\partial V(x(u))}{\partial x(u)}\right).
\label{eq:5}
\end{equation}
We can also arrive at Eq. (4) by substituting Eq. (1) directly into
Newton's equations of motion.  The action formulation is more useful,
however, since it easily generalizes to finite temperature situations where
it is convenient to regard the time variable as a complex quantity \cite
{six}.  Moreover, the time derivative of the action, a quantity that is
easily constructed in practice, provides a convenient measure of the path's
energy.  Specifically, the energy of a classical path, $E_{path}$, can be
expressed as \cite {nine}
\begin{equation}
\frac{\partial S}{\partial t} = - E_{path},
\label{eq:6}
\end{equation}
which from Eqs. (1) and (2) is easily shown to be
\begin{equation}
\frac{\partial S}{\partial t} = - (S/t + 2 \int_0^1 V(x(u)) du).
\label{eq:7}
\end{equation}

As noted above, we have flexibility with respect to the choice of the
functional form used to describe the classical path.  Rather than the
Fourier ansatz in Eq. (1), for example, we could utilize the positions of
the trajectory at intermediate times as the path variables.  Gillilan and
Wilson's recent approach \cite {eight} to the double-ended trajectory
problem
is based on such a ``Trotter-like'' construction.  As with analogous
path-integral applications \cite {six}, the choice of path description is
somewhat arbitrary.  We note, however, that Eq. (1) has a number of
convenient features.  Unlike segmented path constructions, for example, the
ansatz specified by Eq. (1) is continuous at all levels of truncation.
Moreover, Fourier path expansions avoid problems associated with ``stiff''
degrees of freedom that are produced in Trotter constructions of high
order.  The present approach also introduces a convenient set of length
scales into the problem.  In particular, we see from Eq. (4) that the
higher order coefficients (large $k$) tend toward zero (i.e. become
free-particle like) as $k$ is increased for $t$ fixed if the Fourier sine
component, $f_k$, is bounded.  Conversely, at sufficiently long times the
low-order Fourier coefficients tend to lie on the hypersurfaces defined by
$f_k(a) = 0$.  Finally, Eq. (4) makes explicit certain relationships \cite
{ten} that are somewhat less obvious within other formulations .  For
example, if the potential involved is characterized by a strength
parameter, $\epsilon$, and a natural length scale, $\sigma$, then from Eq. (4)
we see that
the combination $\epsilon t^2/m\sigma^2$ is a similarity parameter for the
associated
dynamics.  That is, systems having a common value of this parameter possess
physically similar trajectories.

The equations for the expansion coefficients (Eq. (4)) are self-consistent
in that these coefficients both determine and depend on the force along the
path.  If the potential energy involved is quadratic (or less), then the
equations that determine these coefficients are linear and hence have at
most a single solution.  For more general potentials, however, the
equations are nonlinear and multiple paths are possible.  The determination
of the required double-ended classical paths is thus reduced to solving the
set of self-consistent equations denoted by Eq. (4).

It is useful to recast the classical dynamics problem (the solution of Eq.
(4)) as an equilibrium minimization problem involving a fictitious
``potential'' energy in a higher-dimensional space.  Specifically, defining
the function $\chi(a)$ as
\begin{equation}
\chi(a) = \sum_{k=1} \epsilon^2_k
\left(a_k + \frac{t^2}{m\pi^2k^2}f_k(a)\right)^2 ,
\label{eq:8}
\end{equation}
where the $\epsilon_k$ variables are arbitrary real constants, we see
that the
desired dynamical information (the Fourier coefficients that determine the
classical path(s)) corresponds to the global minima (zeroes) of this
auxiliary ``potential'' energy.  The topology of $\chi(a)$ dictates the
nature of
the dynamical solution: a single zero implies only one double-ended
trajectory exists for the specified conditions whereas multiple zeroes
imply that many such trajectories are present.  When they exist, we are
often interested in devising strategies for moving between regions
of $\chi(a)$
that correspond to different paths \cite {four,five,six,eleven}.
With this in mind,
we note that we have control of the height of ``barriers'' separating such
minima since we are at liberty to vary the constants $\{\epsilon_k\}$
appearing in Eq.(8).

We have utilized traditional minimization \cite {twelve} and annealing
methods
\cite {thirteen} as well as quantum annealing techniques
\cite {fourteen,fifteen,sixteen} to
locate the required minima of Eq. (8).  This latter method locates minima
of specified functions (in this case $\chi(a)$) using ground state diffusion
Monte Carlo techniques.  It is interesting to note that when used in the
present context such techniques amount to a ground state quantum-mechanical
approach to classical dynamics.  Attempting to solve nonlinear equations
through minimization procedures of any type is often ill-advised since it
is difficult to distinguish local and global minima of the objective
function \cite {twelve}.  Here, however, the solution(s) we seek correspond
to
zeroes of the function in question and are thus easily distinguished from
local minima.

\section{Numerical Application and Discussion}
To illustrate the ideas discussed in the previous section, we examine the
dynamics of a model, two-dimensional anharmonic system, the Henon-Heiles
problem [17].  In this model, a particle of unit mass is assumed to move on
the potential energy surface defined by
\begin{equation}
V(x,y) = \frac{1}{2} x^2 + \frac{1}{2} y^2 + xy^2 - \frac{1}{3} x^3 .
\label{eq:9}
\end{equation}
The physical potential is characterized by a shallow minimum $(V(0,0) = 0)$
and three symmetrically located saddle points $(V(1,0) = V(-1/2, \pm
\sqrt{3}/2) =
1/6)$.  Small amplitude vibrational frequencies of note are those for motion
about the origin (two degenerate frequencies equal to unity) and that for
vibrations perpendicular to the dissociation path at the saddle points
(equal to $\sqrt{3}$).

A convenient device for summarizing the dynamical information is a
two-dimensional display of the energies of the various classical paths
(c.f. Eqs. (6) and (7)) plotted as a function of the trajectory duration.
The topology of the resulting energy-time plot reflects the character of
the dynamics and the underlying potential energy surface.  Such figures,
presented below, were constructed pointwise from roughly $10^5$ individual
energy-time values, each obtained using the minimization procedures
discussed in Section II starting from randomly chosen values for the
expansion coefficients.  The number of Fourier coefficients used was
increased in each case until convergence of the resulting plot was
achieved.

Figure (1) is representative energy-time plot for trajectories that begin
and end at a common point, taken here to be the origin, $(x=0,y=0)$.  Such
plots change in detail but not in basic computational complexity as the
initial/final trajectory position is varied.  Each of the curves in Fig.
(1) represents a distinct type of classical trajectory for the model
system.  For example, the $E = 0$ line in Fig. (1) corresponds to
trajectories where the particle remains motionless at the potential
minimum.  Such trivial solutions ($\{a_k\} = 0$, all $k$) are always
possible when the beginning/end point of the path is an extremum of the
potential.

We see that the complexity of Fig. (1) increases with time as new solutions
appear.  Broadly speaking, these new solutions emerge as primary or
secondary branches from the $E = 0$ solution, or as features that descend
from energies above the classical threshold, $E = 1/6$.  This second class of
solutions resides principally (but not exclusively) above the dissociation
threshold.  Figure (1) is a summary of the distinct energies for paths that
satisfy the required double-ended boundary conditions at various times.  It
indicates both the total number of such paths, and the number that fall
within a particular energy range.  The latter information is central in
deciding which paths are relevant to the dynamical process of interest for
finite temperature systems.

To understand the general character of the various features, it is useful
to note that if the potential were purely harmonic, normal mode vibrations
would give rise to a series of vertical lines in Fig. (1) at half-integer
multiples of the natural periods for small amplitude motion.  Such
solutions would have an infinite slope ($dE/dt ->\infty$) since the
period for
harmonic motion is independent of its energy.  Recalling that the normal
mode frequencies for present system are unity, we identify the features
emerging from the zero energy solution at time intervals of $\pi$ as
reminants
of such small amplitude vibrational motion.

Anharmonic effects modify the small amplitude vibrational features and give
rise to the curvature evident in Fig. (1).  In particular, the dissociative
character of the potential makes possible vibrational motion of arbitrary
duration by allowing the system to (essentially) come to rest atop the
dissociation barrier.  The series of curves asymptotically approaching the
dissociation limit, $E = 1/6$, are characteristic of such motion.  Distinct
differences between the low energy behavior of the features occurring at
times equal to odd and even multiples of $\pi$ can be understood by recalling
the half-period for motion that starts at the potential minimum and is
directed at the repulsive (attractive) wall of a typical one-dimensional
molecular potential {\it decreases (increases)} with increasing energy \cite
{eighteen}.  The two branches of the odd-$\pi$  features thus reflect
an unequal
number of inner and outer turning points of the associated anharmonic
motion.  It is important to note that the low-energy slopes of the
vibrational features differ fundamentally from the corresponding harmonic
estimates.  Finally, from their time spacing and from plots of the
associated trajectories, we associate the regular series of features seen
emerging from the dissociative branches of the various vibrational
solutions in Fig. (1) with trajectories that have oscillations
perpendicular to the dissociation channel.

The features that descend from above the dissociation limit ($ E = 1/6$) in
Fig. (1) are classical analogs of resonances.  If properly oriented, such
trajectories can persist for extended periods of time before they find the
dissociative portion of the potential energy surface.  At certain times
(e.g. $t \approx  15$), bound motion emerges from this family of solutions.
Unlike the previous vibrational features, we see in Fig. (1) that resonance
solutions proliferate explosively with increasing time.  Similar behavior
in more complex systems can also be anticipated in the vicinity of
isomerization thresholds.

Figure (2) represents a similar energy-time plot for the Henon-Heiles
system, except that unlike Fig. (1) the initial and final positions for the
trajectories are distinct.  Here the paths begin at the origin, $(x,y) =
(0,0)$, and end at a point beyond the dissociation
barrier, $(x',y') = (3/2,0)$.
 Since the dissociation is an activated process, there is a minimum energy
threshold for such paths and the trivial, zero energy solution is absent.
At very short times, motion across the barrier is essentially ballistic and
the trajectory energy scales as $1/t^2$.  The left-most
branch visible in Fig.
(2) corresponds to such a direct dissociative path.  The second major
branch that descends from $E = 0.5$ near $t = 4$ is a similar dissociative
path, except that here the particle has suffered a single reflection from
the repulsive wall of the potential opposite the dissociation channel.
With increasing time, the energies of these and other more involved
vibrational trajectories decrease toward a value equal to the barrier
height.  Such asymptotic limits thus serve as a measure of the barrier
height for dissociation.  As in Fig. (1), the series of regularly spaced,
near vertical features seen in Fig. (2) are associated with trajectories
that undergo oscillatory motion perpendicular to the dissociation
coordinate.  The ``domain'' structure seen in the energy-time plots suggests
that these and other solutions exhibit a ``non-crossing'' behavior.

\section{Summary}
We have presented methods for the construction of double-ended classical
trajectories.  Based on Fourier path expansions, the present approach
reduces the original dynamical problem to one of finding the minima of a
fictitious, high-dimensional classical potential energy.  Numerical
applications to a model, anharmonic system, the Henon-Heiles potential,
reveal a variety of general energy-time relationships for the double-ended
trajectories of the system.

For purposes of illustration, the present work has focused on a relatively
simple dynamical system.  Although the limits of the approach described
here are not yet completely established, it is appropriate to note that
they extend well beyond the present low-dimensional, model applications.
We have found, for example, that using such approaches we can successfully
construct various double-ended trajectories for simple atomic clusters of
various sizes.  These preliminary studies suggest that many of the generic
features observed in Figs. (1-2) are present in the dynamics of these more
general systems.  Details of these cluster applications will be presented
elsewhere.

\section{Acknowledgments}
The authors would acknowledge support through the National Science
Foundation grant NSF-CHE-9203498.  DLF also wishes to acknowledge the
donors of the Petroleum Research Fund of the American Chemical Society for
partial support of this work.

\newpage
\section*{Figure Captions}

\begin{enumerate}

\item
Figure 1(a)  Energy-time plot for the Henon-Heiles system.  Paths begin and
end at the minimum of the potential and contain up to twelve Fourier
coefficients in their description.

\item
Figure 1(b)  As in Fig. (1a) for longer times to explore the increase in
number and complexity of the $E\geq  1/6$ solutions.

\item
Figure (2)  As in Fig. (1a), except that paths are dissociative in
character ($(x,y) = (0,0), (x',y') = (3/2,0)$).
\end{enumerate}
\end{document}